\documentclass[fleqn,usenatbib,useAMS]{mnras}
\usepackage{graphicx}    
\usepackage{amsmath}    
\usepackage{amssymb}    
\usepackage{multicol}
\usepackage{multirow}
\usepackage{mathtools}
\usepackage{caption}

\usepackage[T1]{fontenc}
\usepackage{ae,aecompl}
\usepackage{newtxtext,newtxmath}
\usepackage{threeparttable,tablefootnote}
\usepackage{url}
\usepackage{hyperref}
\usepackage{placeins}
\usepackage{gensymb}
\usepackage[dvipsnames]{xcolor}
\usepackage{etoolbox}
\usepackage{float}

\makeatletter
\patchcmd\@combinedblfloats{\box\@outputbox}{\unvbox\@outputbox}{}{\errmessage{\noexpand patch failed}}
\makeatother

\numberwithin{equation}{subsection}
\numberwithin{table}{section}
\numberwithin{figure}{section}

\title[]{Strong lensing as a giant telescope to localize the host galaxy of gravitational wave event}
\author[]{Hai Yu$^{1}$\thanks{E-mail:\href{mailto:yu\_hai@sjtu.edu.cn}{yu\_hai@sjtu.edu.cn}}, 
	Pengjie Zhang$^{1,2,3}$,
	Fa-Yin Wang$^{4,5}$\\
	$^{1}$Department of Astronomy, School of Physics and Astronomy, Shanghai Jiao Tong University, Shanghai, 200240, China\\
	$^{2}$Tsung-Dao Lee institute, Shanghai Jiao Tong University, Shanghai, 200240, China\\
	$^{3}$Shanghai Key Laboratory for Particle Physics and Cosmology, Shanghai, 200240, China\\
	$^{4}$School of Astronomy and Space Science, Nanjing University, Nanjing 210093, China\\
	$^{5}$Key Laboratory of Modern Astronomy and Astrophysics (Nanjing University), Ministry of Education, Nanjing 210093, China}

\date{\today}
\pubyear{2020}
\begin{document}
	\label{firstpage}
	\pagerange{\pageref{firstpage}--\pageref{lastpage}}
	\maketitle
	\begin{abstract}
		Standard siren cosmology of gravitational wave (GW) merger events relies on the identification of host galaxies and their redshifts. But this can be highly challenging due to numerous candidates of galaxies in the GW localization area. We point out that the number of candidates can be reduced by orders of magnitude for strongly lensed GW events, due to extra observational constraints. For the next-generation GW detectors like Einstein Telescope (ET), we estimate that this number is usually significantly less than one, as long as the GW localization uncertainty is better than $\sim 10\, \rm deg^2$. This implies that the unique identification of the host galaxy of lensed GW event detected by ET and Cosmic Explorer (CE) is possible. This provides us a promising opportunity to measure the redshift of the GW event and facilitate the standard siren cosmology. We also discuss its potential applications in understanding the evolution process and environment of the GW event.
	\end{abstract}
	
	
	\begingroup
	\let\clearpage\relax
	\endgroup
	\newpage
	
	\section{Introduction}\label{sec:introduction}
	Since the first gravitational wave (GW) event GW150914 caused by binary black holes (BBH) merger was detected by Advanced LIGO \citep{Abbott2016PhRvL.116x1103A}, a dozen of GWs from compact binary mergers have been detected by LIGO and Virgo in their O1 and O2 observations. Among them, GW170817 is the first detected GW event from binary neutron stars (BNS) \citep{Abbott2017PhRvL.119p1101A}. Fortunately, it has electromagnetic (EM) counterpart observations, so we can localize its host galaxy and measure its redshift \citep{Abbott2017ApJ...848L..13A,Abbott2017ApJ...848L..12A}. Recently, a new event, GW190425, was detected as a likely candidate of GW from BNS but without any detection of EM counterpart \citep{Abbott2020ApJ...892L...3A}. These events provide us very good laboratories to test general relativity \citep{Abbott2016PhRvL.116v1101A,Abbott2019arXiv190304467T}. For the next generation GW detectors, such as the Einstein Telescope (ET) \citep{Punturo2010CQGra..27s4002P} and Cosmic Explorer (CE) \citep{Abbott2017CQGra..34d4001A}, they will detect about $10^5$ GW events from BBH, BNS and black halo-neutron star (BH-NS) mergers each year. These GWs can be used as standard sirens to measure cosmological luminosity distance directly \citep{Schutz1986Natur.323..310S}, independent of any astronomical distance ladders and free of any systematic errors associated with them. Therefore, they have many applications in cosmology, such as measuring the Hubble constant \citep{Schutz1986Natur.323..310S,Abbott2017Natur.551...85A}, understanding dark energy \citep{Holz2005ApJ...629...15H,Cutler2009PhRvD..80j4009C,Mendonca2019arXiv190503848M} and constraining the dark matter clustering \citep{Congedo2019PhRvD..99h3526C}.
	
	These applications usually require the measurement of host galaxy redshift.\footnote{Alternatively, some methods are proposed to apply these standard sirens without redshift measurement in cosmology \citep{Namikawa2016PhRvL.116l1302N,Oguri2016PhRvD..93h3511O,Zhang2018arXiv181011915Z,Soares-Santos2019ApJ...876L...7S}. Nevertheless, the cosmological applications will be limited, without the redshift information. } This may be done by follow-ups of EM counterparts \citep{Abbott2017PhRvL.119p1101A,Abbott2017ApJ...848L..13A,Abbott2017ApJ...848L..12A}. \cite{Fan2014ApJ...795...43F} proposed to arrange the strategy of optical follow-up observations by a Bayesian approach. It can improve the probability of finding their EM counterparts. However, this kind of follow-up may only work for nearby GWs. Furthermore, about $90\%$ of the targets of ET will be the GWs from BBH systems which may have no EM counterparts \citep{Biesiada2014JCAP...10..080B}. Another approach is to match known galaxies of pre-existing galaxy surveys to the corresponding GW event according to their properties such as angular positions and distances \citep{Schutz1986Natur.323..310S,Chen2018Natur.562..545C,Fishbach2019ApJ...871L..13F,Soares-Santos2019ApJ...876L...7S}. However, for the majority of GW events at $z\sim 1$ that will be detected by future GW experiments, the number of possible hosts can be orders of magnitude greater than $1$, and the estimated redshift will have large uncertainty. 
	
	Here we propose a new method to identify the host galaxies, although only applicable to strongly lensed GW (SLGW) events. In a strong lensing system, the light rays from a background source will be bent by an intervening galaxy and form multi-images. As an important tool in astronomy, strong lensing systems have been used in many fields in cosmology. It can be used to investigate the structure and evolution of galaxies \citep{Treu2006ApJ...640..662T,Cao2016MNRAS.461.2192C}, the cosmic curvature \citep{Bernstein2006ApJ...637..598B,Rasanen2015PhRvL.115j1301R,Xia2017ApJ...834...75X}, and the dynamical properties of dark energy \citep{Cao2015ApJ...806..185C}. There are two observable consequences in strong lensing system. The first one is the time delay among the different formed images which can be used to constrain the Hubble constant \citep{Wong2019arXiv190704869W}. The other one is that the formed images are brighter than the background source if there is no strong lensing effect which is called magnification. Similarly, a GW signal from a background source is also probably strongly lensed by an intervening galaxy and form an SLGW system although we haven't detected such an event yet. However, some authors have investigated the probability of detecting the SLGW system in the era of next generation GW detectors. They found that we might detect dozens of SLGWs each year due to the high detection rate of GWs in the era of ET and CE \citep{Piorkowska2013JCAP...10..022P,Biesiada2014JCAP...10..080B,Li2018MNRAS.476.2220L}. This promising prediction encourages people to consider the applications of SLGW in astrophysics and cosmology, such as constraining the speed of GW \citep{Fan2017PhRvL.118i1102F,Collett2017PhRvL.118i1101C}, weak equivalence principle \citep{Yu2018EPJC...78..692Y}, modified gravity \citep{Yang2019ApJ...880...50Y}, Hubble constant \citep{Liao2017NatCo...8.1148L}, as well as understanding the wave nature of GW \citep{Hou2019arXiv191102798H}. However, most of these applications still require the redshift measurement of GW. In this case, it is more unlikely to measure the redshift because the event rate of SLGW with EM counterpart detection is tiny. \footnote{The probability of a typical galaxy being strongly lensed by foreground galaxy is only about $10^{-4}$ \citep{Biesiada2014JCAP...10..080B}. Besides, the probability to detect the EM counterpart of GW is also very small since we only have one such event up to date \citep{Abbott2017ApJ...848L..13A,Abbott2017ApJ...848L..12A}.}
	
	Nevertheless, by matching strong lensing systems in pre-existing galaxy surveys such as LSST \citep{LSST2009arXiv0912.0201L},  the host galaxy may be identified unambiguously. Our proposal is based on two assumptions/requirements. First, the host galaxy of an SLGW is also strongly lensed by the same lens galaxy, forming a strong lensing system in optical/near-IR band.  Since the typical Einstein radius and the typical size of a galaxy stellar component are both around one arc-second, this assumption is reasonable. One exception is that the GW progenitor locates so far away from the center of its host galaxy that the host galaxy itself is not strongly lensed. In this case, we will fail to identify SLGW using galaxy surveys. But it does not lead to a wrong assignment to another host galaxy. Second, we assume that the galaxy strong lensing systems can be observed by optical/near-IR galaxy surveys, before the third generation GW experiments. Galaxy surveys such as LSST, WFIRST, and Euclid are powerful for strong lensing search. For example, LSST is expected to discover about $3\times 10^4$ galaxy-galaxy strong lensing systems in its 10-year 20,000 deg$^2$ survey \citep{LSST2009arXiv0912.0201L}.  We also assume that the lenses of these strong lensing systems are identified as well, either by the same surveys or by follow-up surveys. We are then able to model these systems and compare the predicted time delay and magnification of galaxy images with the observed SGLW time delay and magnification. These extra constraints, along with the improved localization of ET and CE (sub-deg$^2$ to dozens of deg$^2$ \citep{Zhao2018PhRvD..97f4031Z}),  can significantly reduce the ambiguity in matching GW events with galaxies. In fact, we find that, for ET and CE,  the mean number of candidates passing all the above constraints is usually significantly smaller than one, implying unambiguous identification of host galaxies with our proposal. 
	
	Our paper is organized as follows. We introduce our method in detail in section \ref{sec:Method}. Then we present our main results and discussion in section \ref{sec:Results_Discussion}. Finally, we give our conclusion in section \ref{sec:Conclusion}. We adopt a flat-$\Lambda$CDM model with $H_0=70\,\rm km/s/Mpc$ and $\Omega_m=0.3$ as our fiducial cosmology.
	
	\section{Method}\label{sec:Method}
	Our method can be summarized by the following steps:
	\begin{itemize}
		\item {\bf Step 1}. Given a SLGW event detected by the next generation GW detectors, we determine its localization $\mathbf{\delta \Omega}$, time delay $\Delta t_{\rm GW}$ and magnification ratio $R_{\rm GW}$. For a conservative localization accuracy $\mathbf{\delta \Omega}=\mathcal{O}(10){\,\rm deg}^2$ of next generation GW experiments \citep{Zhao2018PhRvD..97f4031Z}, the total number of galaxies (potential GW hosts) accessible to LSST and other galaxy surveys is $\mathcal{O}(10^6)$. Therefore we need significantly more efficient criteria for identifying the host. 
		\item {\bf Step 2}.  In the area of $\mathbf{\delta \Omega}$, we search for all strong lensing systems in galaxy surveys, identified by multiple images or giant arcs. The total number of potential hosts then reduces from $\mathcal{O}(10^6)$ to $\mathcal{O}(10^2)$, taking LSST as an example. This is about 4 orders of magnitude reduction because the probability of a galaxy being strongly lensed by foreground galaxy is only about $10^{-4}$ \citep{Biesiada2014JCAP...10..080B}. Nevertheless, we need a further selection process. 
		\item {\bf Step 3}. Within these strong lensing systems, we search for the ones capable of reproducing the GW magnification ratio $R_{\rm GW}$  and time day $\Delta t_{\rm GW}$.  (1) We estimate the magnification ratio $R_{\rm g}$ from the flux of multiple galaxy images, which should be close to $R_{\rm GW}$.  (2) For each system, we model the lens by using the observations of lens galaxy, positions of images, and so on to theoretically estimate the GW  time delay $\Delta t_{\rm th}$. We then search for candidates with $\Delta t_{\rm th}$ and $R_g$ matching $\Delta t_{\rm GW}$ and $R_{\rm GW}$, within reasonable observational and theoretical uncertainties.  This selection criterion is efficient.  We find that the mean number of candidate reduces from $\mathcal{O}(10^2)$ in step 2 to $\mathcal{O}(0.1)$.  Therefore the host galaxy can be uniquely determined. 
	\end{itemize}
	
	The probability of detecting the strongly lensed GWs in the era of ET has been well discussed in literature \citep{Piorkowska2013JCAP...10..022P,Biesiada2014JCAP...10..080B,Li2018MNRAS.476.2220L}. Therefore, we refer readers of interest to these references above. Here we focus on whether we can identify uniquely the host galaxy of an SLGW or not if we detect such event in the future. For a strong lensing system which has two images, the time delay between the arriving times of the two images is
	\begin{equation}\label{eq:timeDelay}
	\Delta t = \frac{d_ld_s(1+z_l)}{cd_{ls}}\Delta \psi(\theta_1, \theta_2).
	\end{equation}
	Here $d_l$ and $d_s$ are angular diameter distances of the lens galaxy and lensed object, respectively. $d_{ls}$ is the angular diameter distance between the lens galaxy and the lensed object. $z_l$ and $c$ are the redshift of the lens galaxy and the speed of light. $\Delta \psi(\theta_1, \theta_2)$ is the difference of the Fermat potentials at the positions of the two formed images $\theta_1$ and $\theta_2$. For the singular isothermal sphere (SIS) model, the Fermat potential difference can be expressed easily as $\Delta\psi(\theta_1, \theta_2)=2\beta\theta_E$ where $\beta$ is the position of the source galaxy. The Einstein radius is $\theta_E=4\pi\frac{\sigma_v^2}{c^2}\frac{d_{ls}}{d_s}$ where $\sigma_v$ is the velocity dispersion of the lens galaxy \citep{Narayan1996astro.ph..6001N}. For a strong lensing system, it requires $\beta<\theta_E$. The positions of the two images should be $\beta\pm\theta_E$ respectively. Another parameter we can measure in the strong lensing system is the magnification ratio between multiple images. For SIS model, the magnification rates of the two images are $\mu_\pm = \frac{\theta_E}{\beta}\pm1$. Therefore, the magnification ratio of the two images is $R = \frac{\theta_E+\beta}{\theta_E-\beta}$. In the case of GW, the observable variable is the strain amplitude. Therefore, the ratio of the two GW signal amplitudes should be $\sqrt{R}$ \citep{Wang1996PhRvL..77.2875W}.\footnote{In the rest of this paper, we only use $R$ for both EM and GW cases.}

\begin{figure}
		\centering
		\includegraphics[width=0.47\textwidth]{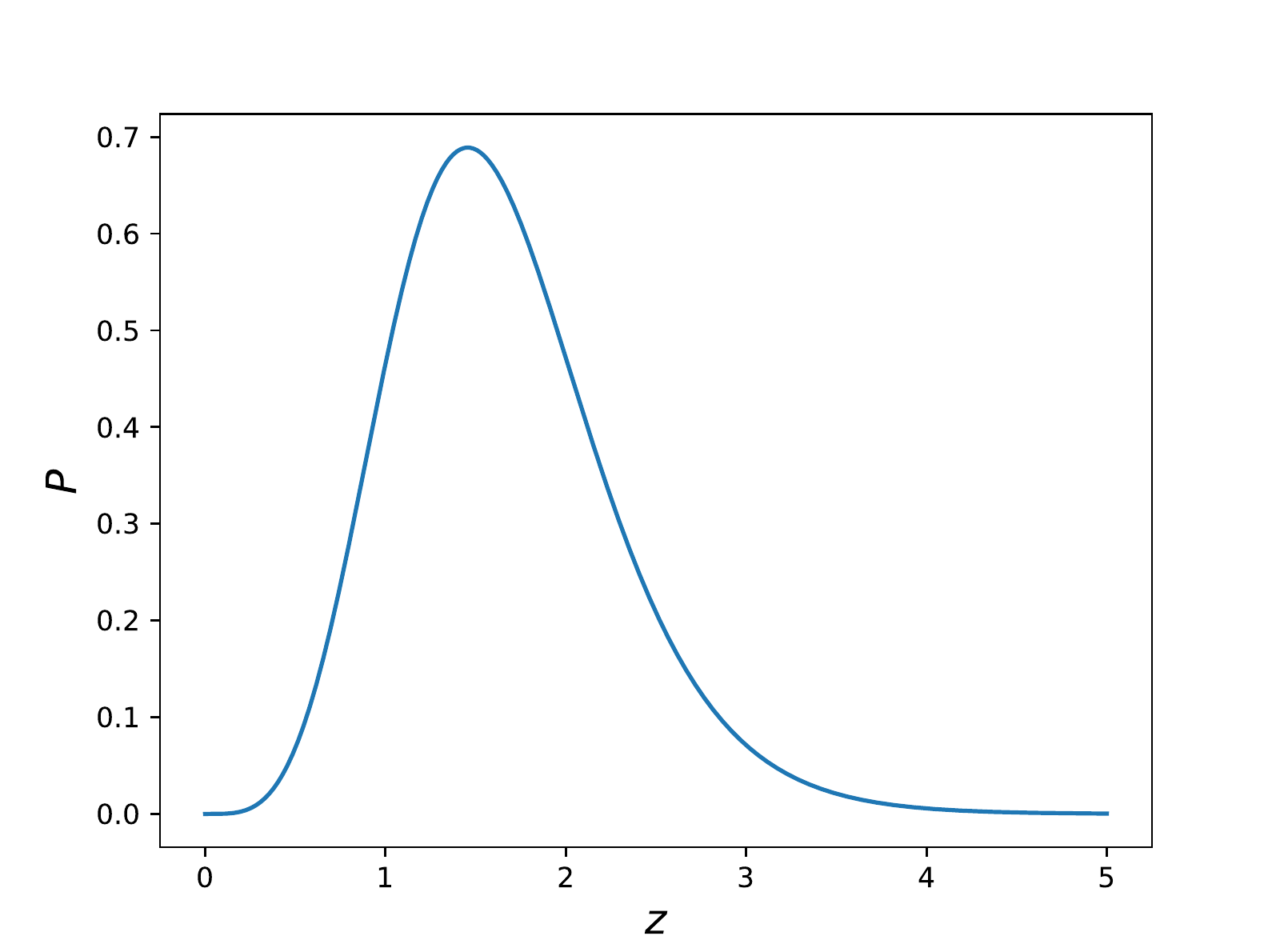}\\
		\caption{The normalized differential distribution of the expected number of the lensed galaxies as a function of redshift.}\label{fig:LensedGalDist}
	\end{figure}

	Now we estimate the average number of strong lensing systems within the region of possible GW sky location ($\mathbf{\delta \Omega}$), and capable of  reproducing the observed $\Delta t_{GW}$ and $R_{GW}$. The probability for an object at redshift $z_s$ being strongly lensed by a foreground galaxy with redshift in $[z_l, z_l+{\rm d}z_l]$ and velocity dispersion in $[\sigma_v, \sigma_v+{\rm d}\sigma_v]$ is
	\begin{equation}\label{eq:dOpticalDepth}
	{\rm d}\tau = \frac{{\rm d}n(\sigma_v,z_l)}{{\rm d}\sigma_v}S_{cr}(\sigma_v,z_l,z_s)\frac{{\rm d}V(z_l)}{{\rm d}z_l}{\rm d}\sigma_v{\rm d}z_l.
	\end{equation}
	Here $n(\sigma_v,z_l)$ is the comoving number density of the lens galaxies per unit velocity dispersion $\sigma_v$ at redshift $z_l$. $S_{cr}(\sigma_v,z_l,z_s)=\pi\theta_E^2$ is the cross-section for a lens of velocity dispersion $\sigma_v$ at $z_l$ and a source at $z_s$.  $V(z_l)$ is the comoving volume within $z_l$. For simplicity and consistency with previous research on detection rate of SLGWs for ET \citep{Piorkowska2013JCAP...10..022P,Biesiada2014JCAP...10..080B,Li2018MNRAS.476.2220L}, we ignore the redshift dependence of $n(\sigma_v,z_l)$, and adopt the following Schechter distribution function
	\begin{equation}\label{eq:SchechterFun}
	\frac{{\rm d}n(\sigma_v,z_l)}{{\rm d}\sigma_v} = n_*\left(\frac{\sigma_v}{\sigma_{v*}}\right)^\alpha \exp\left[-\left(\frac{\sigma_v}{\sigma_{v*}}\right)^\beta\right]\frac{\beta}{\Gamma(\alpha/\beta)}\frac{1}{\sigma_v},
	\end{equation}
	where $(n_*,\sigma_{v*},\alpha,\beta)=(8.0\times10^{-3}\,h^3\,{\rm Mpc^{-3}},161\,{\rm km/s}, 2.32,\\ 2.67)$ comes from the galaxy sample of SDSS DR5 \citep{Choi2007ApJ...658..884C}. $\Gamma(\alpha/\beta)$ is the Gamma function. Integrating eq.(\ref{eq:dOpticalDepth}), one obtains the probability for an object at redshift $z_s$ being strongly lensed. It is \citep{Piorkowska2013JCAP...10..022P}
	\begin{equation}\label{eq:lensingProb}
	\tau(z_s)=\frac{8}{15}\pi^3(1+z_s)^3d_s^3\left(\frac{\sigma_{v*}}{c}\right)^4n_*\frac{\Gamma(\frac{4+\alpha}{\beta})}{\Gamma(\alpha/\beta)}.
	\end{equation}
	To estimate the expected number of strongly lensed galaxies in redshift range $(z_s, z_s+\delta z_s)$, we need the redshift distribution of source galaxies. Here we approximate it by \citep{Holz2005ApJ...629...15H}
	\begin{equation}\label{eq:galaxyDistr}
	\frac{{\rm d}N}{{\rm d}r}=N_0r^a\exp(-(r/r_*)^b),
	\end{equation}
	where $(a,b,r_*)=(1,4,c/H_0)$ and $r$ is the comoving distance \citep{Kaiser1992ApJ...388..272K,Hu1999ApJ...522L..21H}. The normalization factor $N_0$ is chosen to satisfy $\int\frac{{\rm d}N}{{\rm d}r} {\rm d}r=40 \,\rm galaxies\, arcmin^{-2}$, corresponding to several important survey projects in the future like LSST, WFIRST and Euclid \citep{Yao2017JCAP...10..056Y}. Then the expected number of strongly lensed galaxies in redshift range $(z_s, z_s+\delta z_s)$ will be 
	\begin{equation}\label{eq:numberOfLensedGalaxies}
	N_{\rm lg} = \int_{z_s}^{z_s+\delta z_s} \tau(z)\frac{{\rm d}N}{{\rm d}r}\frac{{\rm d}r(z)}{{\rm d}z}{\rm d}z.
	\end{equation}
	The normalized differential distribution of the expected number of the lensed galaxies is shown as Figure \ref{fig:LensedGalDist}. It shows that most of lensed galaxies distribute in the range of redshift $(0.5, 2.5)$ and the distribution peaks at $z\approx1.5$.
	
\begin{figure}
		\centering
		\includegraphics[width=0.47\textwidth]{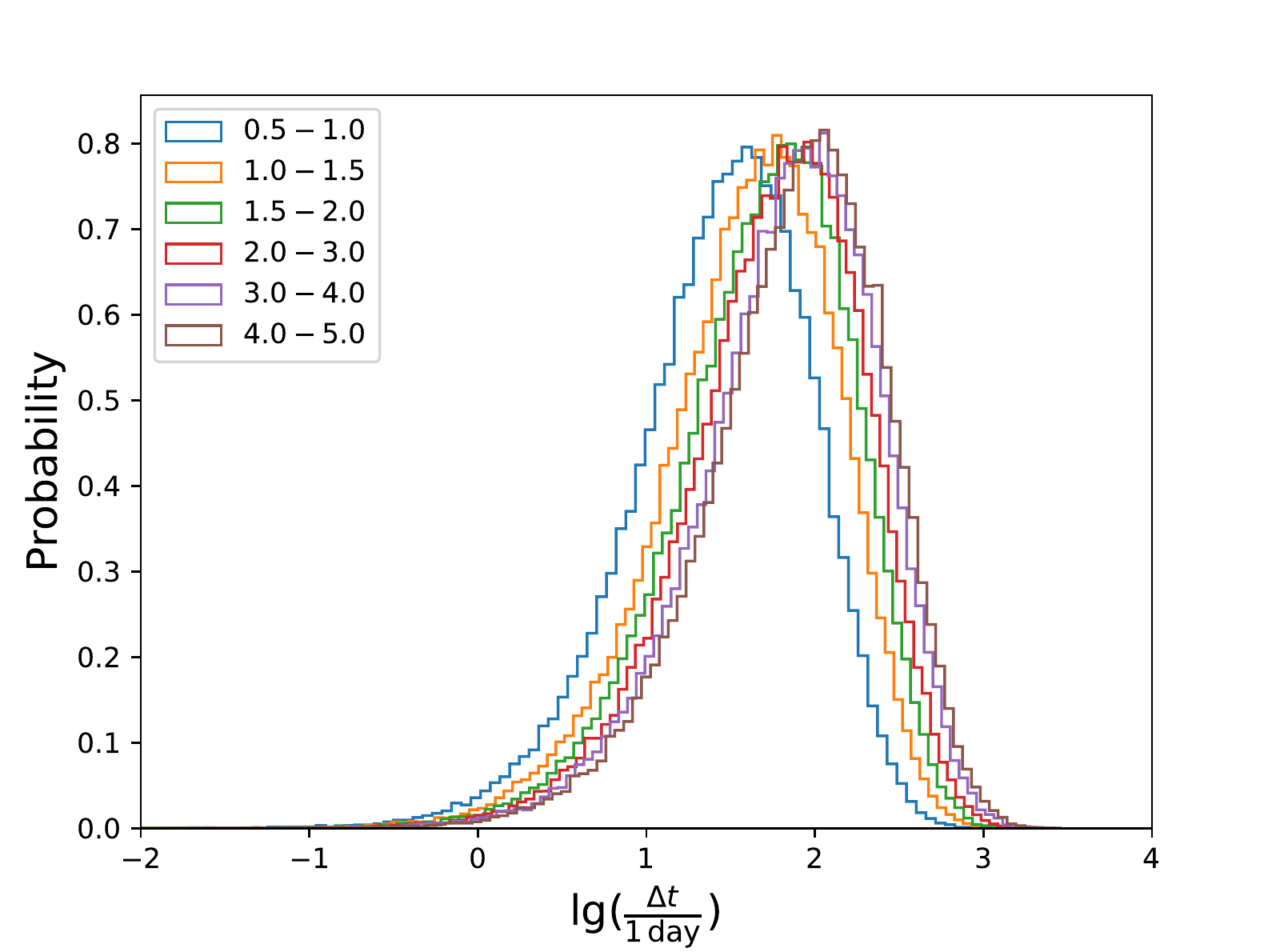}
		\includegraphics[width=0.47\textwidth]{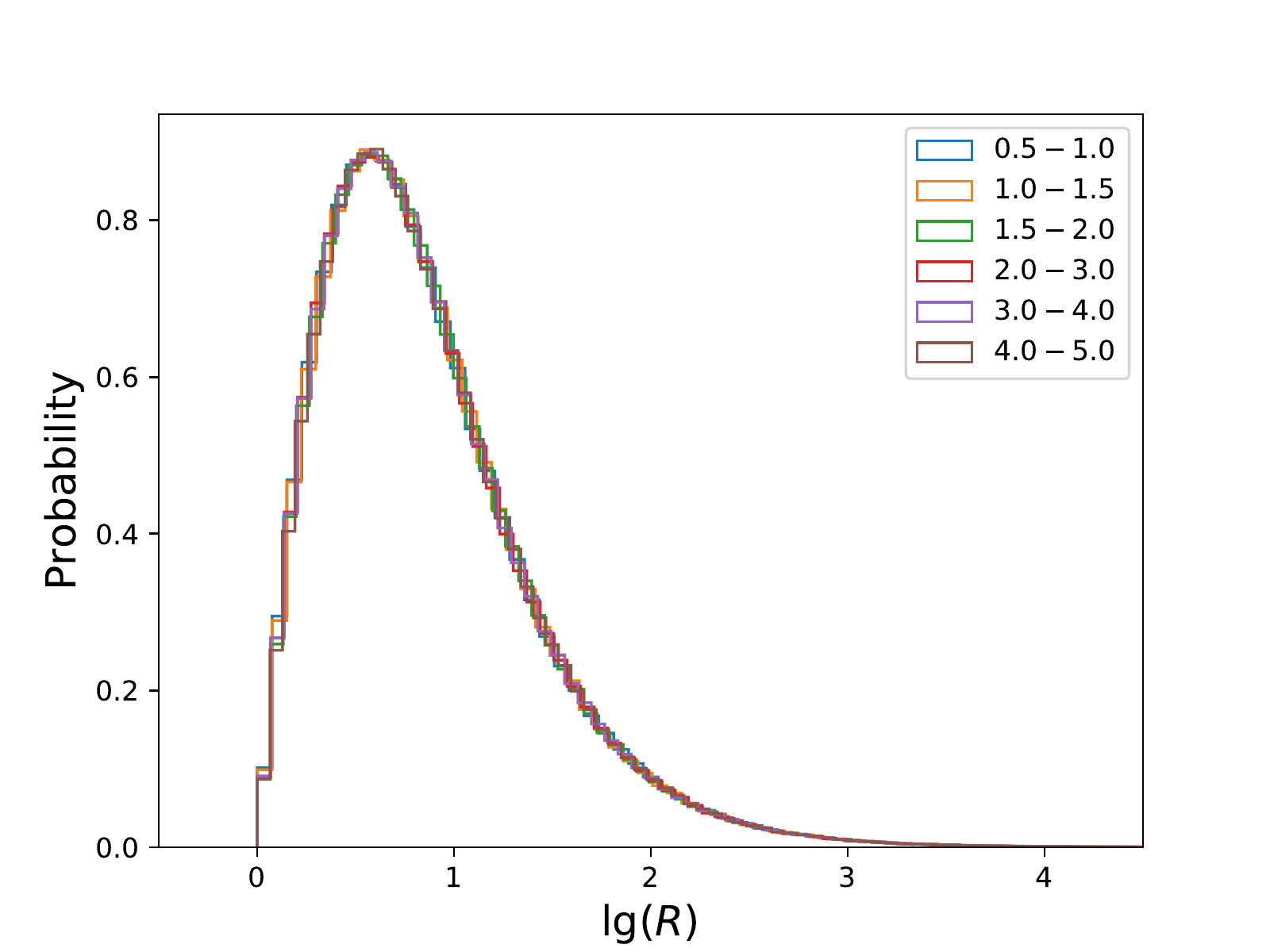}\\
		\caption{Top and bottom panels show the distributions of time delay and magnification ratio of the two formed images in the mock strong lensing systems respectively. The lines with different colors represent the distributions in different redshift ranges of lensed galaxies. Here the distribution of $R$ is independent on the redshift of source because the ratio of $\beta$ and $\theta_E$ in $\mu$ cancels the effect of the $d_s$.}\label{fig:TimeDelayDist_MagnifyRateDist}
	\end{figure}

	Only a fraction of the above strong lensing systems are capable of reproducing the observed GW time delay and magnification. Here we use Monte Carlo simulation to obtain the two distributions and the joint distribution. For each redshift range of lensed object $(z_s, z_s+\delta z_s)$, we generate mock samples of the lens and lensed galaxies by eq. (\ref{eq:SchechterFun}) and (\ref{eq:galaxyDistr}). The angular positions of those galaxies are distributed randomly and uniformly on the sky. Considering the relative positions of the lens and lensed galaxies, we randomly generate $10^5$ mock lensing systems as our mock sample. Then we solve these lensing systems to estimate the distributions of the time delay and magnification ratio (Figure \ref{fig:TimeDelayDist_MagnifyRateDist}). 
	
	\section{Results}\label{sec:Results_Discussion}
	With these distributions, we can estimate the probability of a lensing system with the time delay in arbitrary ranges of time delay, magnification ratio, and source redshift. Multiplying by  $N_{\rm lg}$ in Eq.(\ref{eq:numberOfLensedGalaxies}) and $\delta \Omega$, we will obtain the expected number of candidates in the volume of the above parameter space.
	
\begin{figure}
		\centering 
		\includegraphics[width=0.47\textwidth]{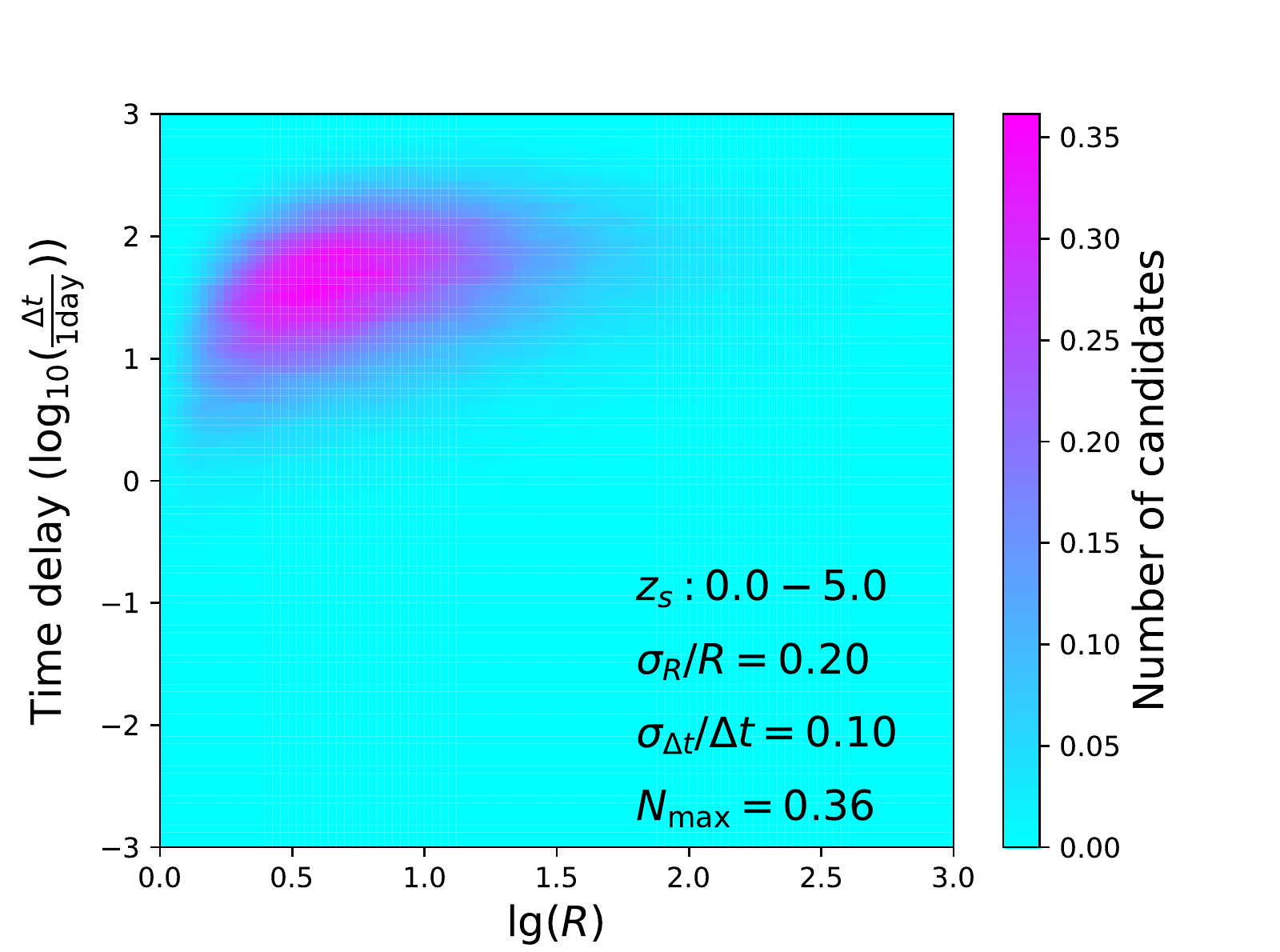}\\
		\caption{The expected number of candidates of the host galaxy as function of $R$ and $\Delta t$ with assumed uncertainties of determined lensing properties.}\label{fig:CandidateDist1}
	\end{figure}

	Now, we estimate the expected number of host galaxies in the sky area $\delta \Omega$ of GW localization, capable of reproducing the observed GW time delay and magnification ratio. ET and CE can localize the GW events within sub-deg$^2$ to dozens of deg$^2$ regions \citep{Zhao2018PhRvD..97f4031Z}, so we adopt $\delta \Omega=10\,\rm deg^2$ in our estimation. Time delay has uncertainties from both the observation of SLGW and the model of lens. However, the uncertainty of the modeling is much larger than that from SLGW observation, so we only consider the model uncertainty. We adopt the fiducial value $\sigma_{\Delta t}/\Delta t=10\%$. If the observed time delay is $\Delta t_{\rm GW}$, any lensing systems with $0.9\Delta t_{\rm GW}\lesssim \Delta t\lesssim 1.1\Delta t_{\rm GW}$ are all likely able to be the GW host. For the magnification ratio, both estimations from SLGW and multiple images in galaxy surveys have observational uncertainties. We take the total uncertainty of $\sigma_{R}/R=20\%$ as the fiducial value. Besides, we assume the redshift of the lensed GW is in a range of $(0, 5)$. Integrating over the possible range of time delay, magnification ratio and $z_s$,  then multiplying by  $N_{\rm lg}$ in Eq.(\ref{eq:numberOfLensedGalaxies}) and $\delta \Omega$, we obtain the expected number $\bar{N}_{\rm candidate}$ of candidates as a function of the observed $R_{\rm GW}$ and $\Delta t_{\rm GW}$ (Figure \ref{fig:CandidateDist1}). We find that, for the majority parameter space of $\Delta t$-$R$,  $N_{\rm candidate}\la 0.1$. The maximum happens for GW events with $\Delta t\simeq 30$ days, and $R\sim 4$. Even for such a case, $N_{\rm candidate}=0.36$, which is still significantly smaller than unity. These results mean that, if we do find a candidate satisfying the above constraints, the likelihood of existing another candidate satisfying the same constraints is rare. Namely, we can identify a unique candidate galaxy hosting the given SLGW, from pre-existing galaxy surveys. The exact value $N_{\rm candidate}$ varies with the assumed lens and source distribution and lens property. Nevertheless, the above numerical result demonstrates that this method is promising. 
	
\begin{figure}
		\centering
		\includegraphics[width=0.47\textwidth]{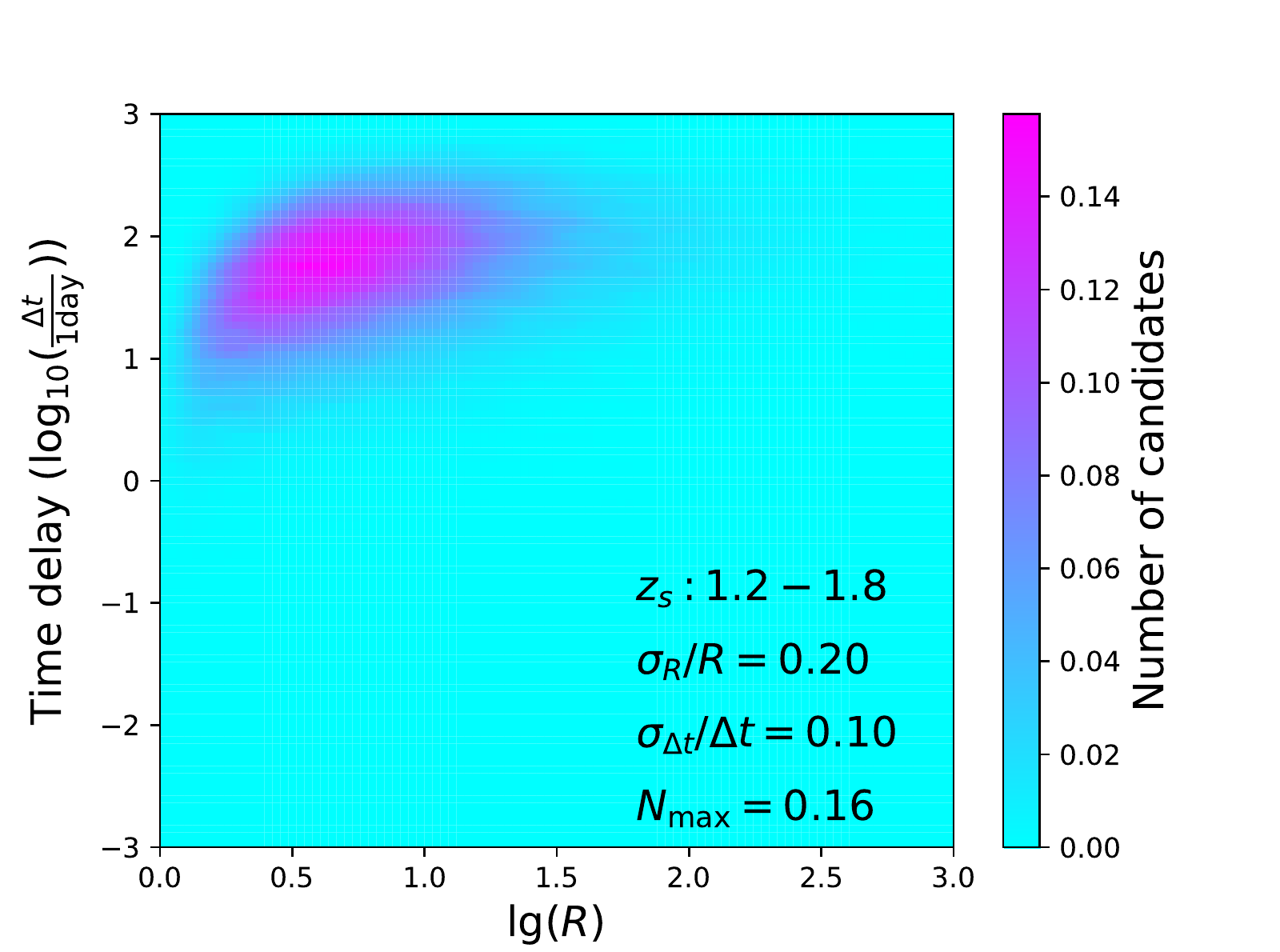}\\
		\caption{Same as Figure \ref{fig:CandidateDist1} but the redshift of lensed galaxy is limited in range of $(1.2, 1.8)$.}\label{fig:CandidateDist2}
	\end{figure}

	The above results adopt the fiducial $\delta \Omega=10 {\rm deg}^2$. Since $\bar{N}_{\rm candidate}\propto \delta \Omega$, for SLGWs with $\delta \Omega \leq 10 {\rm deg}^2$,  their host galaxies can be uniquely identified directly with the method proposed above. However for SLGWs with $\delta \Omega \gg 10 {\rm deg}^2$, we need extra constraint, since $\bar{N}_{\rm candidate}$ may be close to, or bigger than unity. This may come from the source redshift/distance information, which is not used in the previous estimation.  Unlike the case of unlensed GW for which cosmological distance can be determined to good accuracy,  we are not able to accurately determine the luminosity distance directly from the lensed GW signals because of the magnification effect. Nevertheless, we may still be able to obtain some useful constraint on the (true) distance and therefore the redshift $z_s$. For example,  for SIS model, strong lensing events have $\mu_+ = \frac{2R}{R-1}$ and $\mu_- = \frac{2}{R-1}$.  Since $R$ is observable, $\mu_{\pm}$ are observable. We are then able to estimate the true GW amplitude, and therefore the true luminosity distance and redshift.  This estimation is of course only approximate since it strongly relies on the SIS assumption. Nevertheless, it will put a useful constraint on $z_s$, and therefore reduces the number of possible GW host galaxy candidates. This extra information on $z_s$ is most efficient when the constrained $z_s$ range avoids the peak of the strong lensed galaxy redshift distribution, which is around $z_s=1.5$ (Fig. \ref{fig:LensedGalDist}). To be conservative, we consider the opposite case and assume $z_s\in (1.2,1.8)$. The new result of $\bar{N}_{\rm candiate}$ is shown in Figure \ref{fig:CandidateDist2}. Even for this less-efficient scenario, the extra source redshift information is still significantly useful. It reduces $\bar{N}_{\rm candiate}$ by a factor of $2$ and the maximum of $\bar{N}_{\rm candiate}$ is $0.16$. This means that,  even for $\delta \Omega\sim 30\,\rm deg^2$ and the most challenging configuration of $\Delta t$-$R$, it is highly likely that we can identify the GW host galaxy unambiguously. 
	
	The results rely on the assumptions of $\sigma_{\Delta t}$ and $\sigma_{R}$ too. Since the dependences of the probability distribution on $\sigma_{\Delta t}$ and $\sigma_{R}$ are reasonably smooth and $\bar{N}_{\rm candidate}$ is an integral over relatively narrow range of $\Delta t$ and $R$, we expect that $\bar{N}_{\rm candidate}\propto \sigma_{\Delta t}$, and $\bar{N}_{\rm candidate}\propto \sigma_{R}$. The time delay measurement from lensed GW signals can be quite accurate, with a negligible contribution to $\sigma_{\Delta t}$. The magnification ratio measurement error by aLIGO and Virgo can be 30\% at present \citep{Broadhurst2019arXiv190103190B}. ET/CE is about 10 times more sensitive, so we expect much smaller errors. One the other hand, the best uncertainty of the reconstructed Fermat potential of the lens galaxy at present is about 3\% for lensed quasar systems based on the current lensing project H0LiCOW \citep{Suyu2017MNRAS.468.2590S}. And \cite{Liao2017NatCo...8.1148L} claimed that the lens model of GW and host galaxy system would be more precise and accurate. However, this precise and accurate lens model heavily relies on the high resolution imaging of the host galaxy and spectroscopic observations of stellar kinematics of the lens galaxy which we don't expect for every strong lensing system. Therefore, we adopt a more feasible choice of fiducial value $\sigma_{\Delta t}/\Delta t=0.1$ and $\sigma_{R}/R=0.2$. Nevertheless, we also consider the case that $\sigma_{R}/R\in [0.1, 0.4]$ and $\sigma_{\Delta t}/\Delta t\in [0.05,0.2]$ to see their potential effect on our results.
	
\begin{figure}
		\centering
		\includegraphics[width=0.47\textwidth]{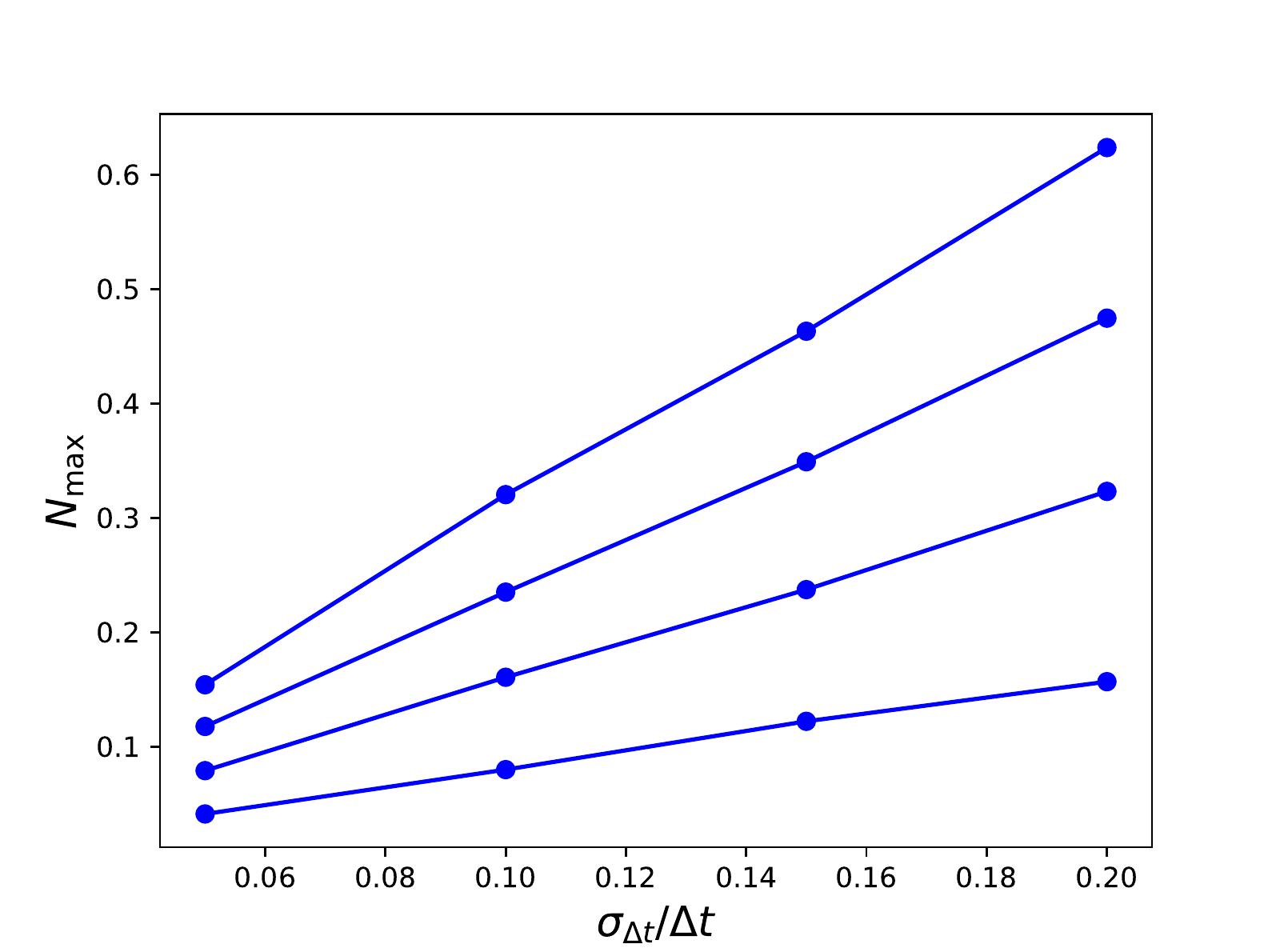}\\
		\caption{The $N_{\rm max}$ as functions of $\sigma_{\Delta t}/\Delta t$ for different values of $\sigma_{R}/R$. The values of $\sigma_{R}/R$ for the lines from bottom to top is $0.1$, $0.2$, $0.3$ and $0.4$, respectively.}\label{fig:UncertaintyFunction}
	\end{figure}

	For most parameter space of $\Delta t_{\rm GW}$-$R_{\rm GW}$,  larger $\sigma_{\rm \Delta t}$ and/or $\sigma_R$ is not able to alter the $\bar{N}_{\rm candidate}\ll 1$ result.  They may only have significant impact in the $\Delta t_{\rm GW}$-$R_{\rm GW}$  parameter space around the peak of $\bar{N}_{\rm candidate}$.  Therefore, we will focus on such case. Assuming the redshift of GW being in the range of $(1.2, 1.8)$ and $\delta \Omega=10\,\rm deg^2$, we estimate the maximum $\bar{N}_{\rm candidate}$  as a function of $\sigma_{\Delta t}/\Delta t$ and $\sigma_{R}/R$ (Figure \ref{fig:UncertaintyFunction}).  Our numerical evaluation confirms our expectation that $\bar{N}_{\rm candidate}\propto \sigma_{\Delta t}$, and $\bar{N}_{\rm candidate}\propto \sigma_R$.  We find $N_{\rm max}\lesssim 0.6$ when $\sigma_{\Delta t}/\Delta t<10\%$ and $\sigma_{R}/R<40\%$. This means that, unless errors in $\Delta t$ and $R$ are much larger than what we have considered, or $\delta \Omega$ is much larger than $10\,{\rm deg}^2$, unique identification of GW host galaxies by our method is feasible. Even if errors in $\Delta t$, $R$ and localization area are much higher than we adopt, the failure of unique identification may only happen when the SLGW $\Delta t_{\rm GW}$ and $R$ locate around the peak region (Fig. \ref{fig:CandidateDist1} \& \ref{fig:CandidateDist2}).

	\section{Discussions and Conclusion}\label{sec:Conclusion}
	We have proposed a method of identifying the host galaxy of SLGW in galaxy surveys. Under simplified conditions, we quantify its performance and demonstrate its promising potential. There are many further complexities that we need to take into account to make a more robust investigation. One is about the lens and source galaxy distribution. For example, we have assumed the number density of the lens galaxies is redshift independent, which is at most an approximation.  We also need to take other observational selection effects into account. GW experiments detect SLGWs of different $\Delta t$ and $R$ with different efficiency and different measurement errors. Galaxy surveys have independent efficiency of detecting strong lensing systems, varying with $\Delta t$, $R$, and other factors. The estimation of $\bar{N}_{\rm candidate}$ should take these extra dependencies into account.  Our estimation presented in this paper does not consider these selection effects and implicitly assumes that galaxy surveys will detect all strong lensing systems. Therefore $\bar{N}_{\rm candiate}$ shown in this paper can only be considered as the upper limit, since galaxy surveys may only detect a fraction of them. This further reduces the confusion rate. Namely, the chance of existing two or more candidates is further reduced. 
	
	On the other hand,  we are not able to reliably quantify the success rate of our method, namely the fraction of SLGWs with successful host galaxy identification.  That means our method may fail to identify the GW host galaxies, simply because that the galaxy surveys fail to detect the associated strong lensing system, either due to the magnitude limit, seeing condition/image separation, or other issues. Quantifying this success rate is beyond the scope of the current paper, and we leave it for future investigation. Furthermore, there are other complexities further reducing the success rate. The first one is that some host galaxies of GWs may be too faint to be detected in future galaxy survey projects. Considering the magnification effect of strong lensing, those host galaxies must be much fainter. Therefore, if there are a part of host galaxies that can not be detected, this means some host galaxies of the GWs are faint. This can constrain the environment of GWs. The second one is the case that the lensed GW lies at the edge of its host galaxy. Then the host galaxy may be out of the Einstein radius of the lens galaxy because of the size of a typical galaxy is comparable to the Einstein radius. In this case, we are unable to find the strong lensing system for the host galaxy. However, we can still search for all the pairs of the galaxies that may host a GW which can form an SLGW system with the same properties as the lensed GW. The expected number of the pairs would not be much larger than the expected number of the candidates of the host galaxy estimated above. This kind of event can help us understand the evolution process of the binary systems. The third one is that we assume the number density of the lens galaxies is constant for different redshift. A much more detailed assumption of this distribution may lead to a modification, but insignificant to alter our major results. The last one is that some lensed host galaxies may not be identified in the future optical galaxy survey projects. For example, LSST is expected to discover about $10^4$ galaxy-galaxy strong lenses in its 10-year 20,000 deg$^2$ survey \citep{LSST2009arXiv0912.0201L}. However, since the ET and CE will not begin operation before the mid of the 2030s \citep{Maggiore2019arXiv191202622M}, we still have time to develop a more advanced algorithm and powerful survey projects. Besides, even though only 10\% strong lensing systems and host galaxies can be identified, we may still determine about 10 GW's host galaxies each year which will contribute a lot to the GW astrophysics and cosmology. \footnote{According to our estimation, there will be about 20 strong lensing galaxy systems in 1 deg$^2$. A LSST-like galaxy survey project can identify 1 or 2 of them which leads to a probability of about 10\%.}

	Gravitational waves are expected to be a powerful tool for many fields of astronomy, such as the physics of neutron stars, black holes, and cosmology. To make full advantage of GWs, it would be better to localize them accurately, and then we can know their host galaxies, redshifts, and so on. However, the poor localization of GWs makes it difficult to identify their host galaxies directly. Fortunately, in the era of ET, hundreds of thousands of GWs will be detected each year and dozens of them will be lensed by intervening galaxies. These lensed GWs provide us an opportunity to identify their host galaxies. In this work, we present a new method to identify the host galaxy of the SLGW event, by using strong gravitational lensing as a giant telescope.

	We summarize our major findings here.  We estimate the expected number of galaxy-galaxy strong lensing systems capable of generating the observed strong lensing GW event, under certain reasonable assumptions. Figure \ref{fig:CandidateDist1} and \ref{fig:CandidateDist2} show our results. For Figure \ref{fig:CandidateDist1}, we only use magnification ratio $R$ and time delay $\Delta t$ with relative uncertainties 20\% and 10\%. We assume the precision of the localization of the lensed GW is $\delta \Omega=10\,\rm deg^2$, and its redshift should be in the range of $(0, 5)$. In this case, the maximum expected number of candidates is about 0.36. It means that we can identify a unique galaxy-galaxy strong lensing system responsible for the lensed GW. Therefore, the lensed galaxy in the galaxy-galaxy strong lensing system should be the host galaxy of the lensed GW. Furthermore, analyzing the observations of lensed GW and galaxy simultaneously, we can make use of the luminosity distance (or redshift) information. Figure \ref{fig:CandidateDist2} shows that this extra information is highly useful, and the expected number of candidates can be reduced significantly. Therefore, we may identify the host galaxy of the lensed GW even if its localization uncertainty is as large as 30 deg$^2$.  We also discussed several other factors that may affect our results (e.g. Fig. \ref{fig:UncertaintyFunction}).  
	
	\section*{Acknowledgments}
	We thank Xi-Long Fan and Ji Yao for useful discussion. HY thanks for the useful discussion in HOUYI Workshop. HY is supported by Initiative Postdocs Supporting Program (No. BX20190206), Project funded by China Postdoctoral Science Foundation (No. 2019M660085), and Super Postdoc Project of Shanghai City. PJZ is supported by the National Science Foundation of China (11621303, 11653003). FYW is supported by the National Science Foundation of China (U1831207).
	
	\bibliographystyle{mnras}
	\bibliography{ref}
	
\end{document}